\documentclass[eat,twocolumn]{jmlr}

\usepackage{multirow}

\usepackage{longtable}

\theorembodyfont{\upshape}
\theoremheaderfont{\scshape}
\theorempostheader{:}
\theoremsep{\newline}

\jmlrvolume{}
\firstpageno{1}

\jmlryear{2022}
\jmlrworkshop{Machine Learning for Health (ML4H) 2022}

\author{\Name{Changye Li} \Email{lixx3013@umn.edu}\\
  \addr Institute of Health Informatics, University of Minnesota
  \AND
  \Name{Trevor Cohen} \Email{cohenta@uw.edu}\\
  \addr Biomedical Informatics and Medical Education, University of Washington
  \AND
  \Name{Serguei Pakhomov} \Email{pakh0002@umn.edu}\\
  \addr Department of Pharmaceutical Care \& Health Systems, University of Minnesota}

\title[The Far Side of Failure]{The Far Side of Failure: Investigating the Impact of Speech Recognition Errors on Subsequent Dementia Classification}

\begin{document}

\maketitle

\begin{abstract}
Linguistic anomalies detectable in spontaneous speech have shown promise for various clinical applications including screening for dementia and other forms of cognitive impairment. The feasibility of deploying automated tools that can classify language samples obtained from speech in large-scale clinical settings depends on the ability to capture and automatically transcribe the speech for subsequent analysis. However, the impressive performance of self-supervised learning (SSL) automatic speech recognition (ASR) models with curated speech data is not apparent with challenging speech samples from clinical settings. One of the key questions for successfully applying ASR models for clinical applications is whether imperfect transcripts they generate provide sufficient information for downstream tasks to operate at an acceptable level of accuracy. In this study, we examine the relationship between the errors produced by several deep learning ASR systems and their impact on the downstream task of dementia classification. One of our key findings is that, paradoxically, ASR systems with relatively high error rates can produce transcripts that result in better downstream classification accuracy than classification based on verbatim transcripts.  

\end{abstract}

\section{Introduction}
\label{intro}

The number of people living with Alzheimer's Dementia (AD) is rapidly growing \citep{alzheimer20182018} and it is estimated that more than half of people living with dementia have not been diagnosed \citep{doi:10.1080/13607863.2011.596805}. Lack of diagnosis leads to unnecessary negative effects on dementia patients and their caregivers, such as suboptimal care planning \citep{https://doi.org/10.1111/psyg.12095}. Simple, timely, cost-effective, and reliable cognitive tests are needed by physicians, researchers, and caregivers \citep{fox2013pros, iliffe2003sooner}. The machine learning community have developed methods to automatically characterize cognitive changes caused by AD using speech and language features (for a review, see \citet{martinez2021ten}). Automatic speech recognition (ASR) models can be used to generate transcripts automatically, and the generated transcripts have shown to be a good source of features for dementia detection models \citep{weiner17_interspeech}. However, the errors (quantified as word error rate (WER) and character error rate (CER)) generated by ASR systems can lower a predictive model's discrimination accuracy in identifying dementia from audio samples \citep{zhou16_interspeech}. Achieving single-digit WERs/CERs on open-domain datasets, Wav2Vec2 \citep{NEURIPS2020_92d1e1eb} and HuBERT \citep{Hsu2021HuBERTSS} are two of the most influential pre-trained self-supervised learning (SSL) ASR models. Wav2Vec2 masks input frames with convolutional features and utilizes contrastive predictive coding (CPC) \citep{oord2018representation} with Transformer \citep{NIPS2017_3f5ee243} encoder blocks. HuBERT, on the other hand, takes the continuous input waveforms and masks input frames with k-means units trained on Mel-frequency cepstral coefficients (MFCC) as features, then pre-trains with pseudo-labeling with Transformer encoder blocks. However, much higher WERs have been reported with this state-of-the-art (SOTA) ASR system for spontaneous speech and there has also been scant work on investigating ASR performance on speech from symptomatic participants captured ``in the wild'' \citep{XU2022103998, balagopalan-etal-2020-impact, Sadeghian2021, DBLP:journals/corr/abs-2110-15704}.

In this paper, we investigate the impact of ASR errors present in imperfect transcripts of speech produced on a picture description task by patients with dementia and healthy controls on the accuracy of subsequent automatic categorization of these transcripts by neural classification models. The current SOTA results on this classification task are obtained using verbatim transcripts manually transcribed by trained human transcriptionists. Theoretically, using imperfect ASR-generated transcripts should result in worse than SOTA classification performance; however, it is currently not known if there is a point at which imperfect ASR performance can be good enough to produce acceptable classification results close to SOTA. Finding the answer to this question will tell us which ASR systems can be readily used in conjunction with neural classification models of dementia to engineer systems that do not rely on manual transcription input - a rate-limiting factor for deployment of such systems in real clinical scenarios. The code for this work is publicly available on GitHub\footnote{\url{https://github.com/LinguisticAnomalies/paradox-asr}}.

\begin{table*}[htbp]
\small
\centering
\caption{Dataset description. Age of the WLS dataset is calculated based on 2011 visit when the ``Cookie Theft'' was administered.}
\begin{tabular}{|ll|ll|ll|}
\hline
\multicolumn{2}{|l|}{\multirow{2}{*}{\textbf{Characteristics}}} & \multicolumn{2}{l|}{\textbf{ADReSS}}    & \multicolumn{2}{l|}{\textbf{WLS}}                            \\ \cline{3-6} 
\multicolumn{2}{|l|}{}& \multicolumn{1}{l|}{Train} & Test & \multicolumn{1}{l|}{Train} & \multicolumn{1}{l|}{Test} \\ \hline
\multicolumn{2}{|l|}{Age, mean (SD)} & \multicolumn{1}{l|}{65.6 (6.5)} & 65.6 (7.0) & \multicolumn{1}{l|}{70.2 (3.2)} & \multicolumn{1}{l|}{70.3 (3.6)}\\ \hline
\multicolumn{1}{|l|}{\multirow{2}{*}{Gender, n (\%)}} & Male & \multicolumn{1}{l|}{48 (44)} & 22 (46) & \multicolumn{1}{l|}{62 (53)} & \multicolumn{1}{l|}{38 (52)}\\ \cline{2-6} 
\multicolumn{1}{|l|}{}  & Female & \multicolumn{1}{l|}{60 (56)} & 26 (54) & \multicolumn{1}{l|}{52 (47)} & \multicolumn{1}{l|}{35 (48)}\\ \hline
\multicolumn{2}{|l|}{Education, mean (SD)} & \multicolumn{1}{l|}{13.1 (3)} &12.9 (2.0)  & \multicolumn{1}{l|}{13.3 (2.1)} & \multicolumn{1}{l|}{13.1 (3.7)}\\ \hline
\multicolumn{2}{|l|}{MMSE, mean (SD)} & \multicolumn{1}{l|}{24.1 (5.9)} &24.9 (5.3) & \multicolumn{1}{l|}{NA} & \multicolumn{1}{l|}{NA}\\ \hline
\multicolumn{2}{|l|}{Number of utterance-level transcripts}  & \multicolumn{1}{l|}{1,414} & 567 & \multicolumn{1}{l|}{1,701} & \multicolumn{1}{l|}{1,146} \\ \hline
\end{tabular}
\label{tab:my-data}
\end{table*}

\section{Methods}
\label{methods}


We used two datasets, a) AD Recognition through Spontaneous Speech (ADReSS) \citep{bib:LuzHaiderEtAl20ADReSS}, and data from the b) Wisconsin Longitudinal Study (WLS) \citep{herd2014cohort}. Transcripts from both datasets were annotated in CHAT format with the Computerized Language Analysis (CLAN) manual annotation system \citep{MacWhinney2000TheCP}\footnote{For more details, please refer to \url{https://talkbank.org/manuals/CHAT.pdf}}. Dataset characteristics are provided in Table~\ref{tab:my-data}. Both datasets contain responses to a widely-used diagnostic task, the ``Cookie Theft'' picture description (Figure~\ref{fig:cookie-theft} in Appendix~\ref{apd}) from the Boston Diagnostic Aphasia Examination \citep{goodglass1983boston}. Participants were asked to describe everything they see going on in the picture, and their responses were audio recorded and then manually transcribed verbatim. We performed basic pre-processing using TRESTLE (\textbf{T}oolkit for \textbf{R}eproducible \textbf{E}xecution of \textbf{S}peech \textbf{T}ext and \textbf{L}anguage \textbf{E}xperiments)\footnote{\url{https://github.com/LinguisticAnomalies/harmonized-toolkit}} to remove artifacts such as speech and non-speech event descriptions (i.e., ``overlap'', ``clear throat''), unintelligible words, and speech that did not belong to participants (i.e., speech of the examiner) from verbatim transcripts, to resample the audio recordings to 16 kHz, and to partition the recordings into utterance-level chunks using time-stamps from the verbatim transcripts. Dataset characteristics are provided in Table~\ref{tab:my-data}.

\paragraph{ADReSS}is matched on age and gender, and a balanced subset of the Pitt corpus in Dementia Bank (DB) \citep{becker1994natural} including the Mini-Mental State Examination (MMSE) scores \citep{folstein1975mini} as neurological test results. The ADReSS dataset contains a total of 156 samples (78 with dementia and 78 controls) split into training and testing subsets, resulting in 1,414 and 567 utterance-level transcripts in the training and test split respectively. 

\paragraph{WLS}is a longitudinal study of a random sample of 10,317 men and women who graduated from Wisconsin high schools in 1957, where the participants were interviewed up to six times between 1957 and 2011. Cognitive evaluations and the ``Cookie Theft'' picture description task were introduced to the later rounds of interview. All of the participants in the WLS were presumably cognitively healthy when entering the study. Some may have developed dementia in later years but the neurological diagnostic result is not currently publicly available. We restricted the original WLS dataset to 87 female and 100 male participants, who a) agreed to participate in the ``Cookie Theft'' picture description task and b) had recording samples with fully aligned verbatim transcripts with the CHAT protocol. We selected the first 52 female and 62 male participants and their output as the training set, and the remaining data as the evaluation set for the WLS dataset, resulting in 1,701 and 1,146 utterance-level transcripts, respectively.

We examined a number of Wav2Vec2 and HuBERT model variations\footnote{All models are available on the HuggingFace model repository: \url{https://huggingface.co/}} that were pre-trained on 16 kHz audio and paired with processors.\footnote{The details of models are listed in Table~\ref{tab:models} in Appendix~\ref{apd}.} Specifically, {Wav2Vec2-base-960h}, {Wav2Vec2-large-960h} and {hubert-large-ls960-ft} were pre-trained and fine-tuned on 960 hours of LibriSpeech \citep{7178964}; {Wav2Vec2-large-960h-lv60} was pre-trained and fine-tuned on 960 hours of Libri-Light \citep{librilight} and LibriSpeech, and {Wav2Vec2-large-960h-lv60-self} was pre-trained and fine-tuned on the exactly same dataset but with a self training objective \citep{Xu2021SelfTrainingAP}.

\begin{figure}
	\hskip-2.5cm\begin{minipage}{0.8\textwidth}
		\centering
		\includegraphics[scale=0.3]{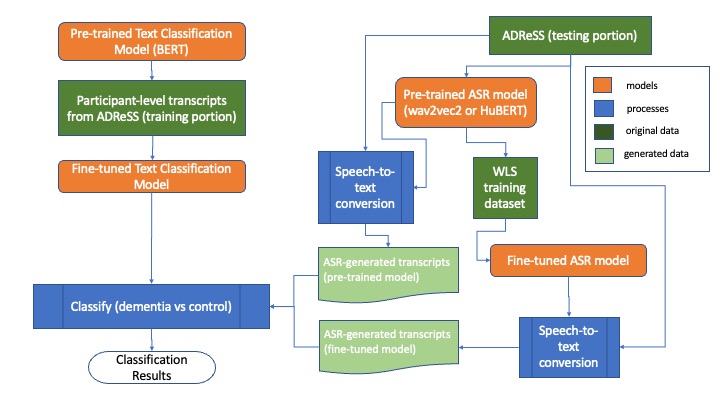}
		\caption{The overview of experimental design}
		\label{fig:overview}
	\end{minipage}
\end{figure}

We evaluated Wav2Vec2 and HuBERT model variations in two ways: a) ``out-of-the-box'': we forwarded the audio samples from the ADReSS test set through these models to generate the corresponding utterance-level transcripts; b) with fine-tuning: we fine-tuned the ASR models with the WLS training set using a batch size of 4 over 10 epochs, and then forwarded the audio samples from the ADReSS test set to the fine-tuned ASR models to generate the transcripts. To assess the classification performance on ASR-generated transcripts, we determined the baseline model performance by following the SOTA approach of identifying dementia using verbatim transcripts. This approach involves fine-tuning a BERT model with the verbatim transcripts from the ADReSS training set then evaluating it on the ADReSS test set \citep{Balagopalan2020ToBO}, achieving an accuracy of 0.826 and an area under the receiver-operator curve (AUC) of 0.873. We evaluated the performance of ASR models using WER/CER (Equation~\ref{eq:wer} in Appendix~\ref{apd}) on the utterance-level transcripts. To perform the evaluation on ASR-generated transcripts, we first merged the utterance-level ASR generated output into participant-level transcripts, and then forwarded the participant-level transcripts through the BERT model fine-tuned on the verbatim transcripts from the ADReSS training set, and compared the classification accuracy and AUC with the baseline performance. The overview of our experimental design is shown in Figure~\ref{fig:overview}.

\section{Results}
\label{results}

\paragraph{Fine-tuning on an additional ``Cookie Theft'' dataset improved the performance of ASR models. But ASR models still struggled with the challenging spontaneous speech.}We observed that pre-trained ASR models performed poorly on the ADReSS and WLS test set, as shown in Table~\ref{tab:res}. The average WER/CER reached 0.466/0.277 on the ADReSS test set, 0.427/0.318 on the WLS test set. Fine-tuning ASR models led to lower WER/CER on the ADReSS and WLS test set, resulting in average WER/CER at 0.383/0.253, and 0.306/0.185, respectively. 

\begin{table*}[htbp]
\centering
\small
\caption{WER/CER performance of pre-trained and fine-tuned ASR models and the subsequent classification performance of accuracy (ACC) and AUC using ASR-generated transcripts evaluated with a BERT model fine-tuned on the verbatim transcripts. The performance of ASR models was evaluated on the utterance-level audio segments, and the subsequent classification performance was evaluated on the participant-level ASR-generated transcripts on the ADReSS test set.}
\begin{tabular}{|l|llll|llll|}
\hline
\multirow{2}{*}{\textbf{Model}} & \multicolumn{2}{p{2.5cm}|}{\textbf{Pre-trained ASR models}} & \multicolumn{2}{p{2.5cm}|}{\textbf{Classification performance}} & \multicolumn{2}{p{2.5cm}|}{\textbf{Fine-tuned ASR models}} & \multicolumn{2}{p{2.5cm}|}{\textbf{Classification performance}} \\ \cline{2-9} 
& \multicolumn{1}{l|}{WER} & \multicolumn{1}{l|}{CER} & \multicolumn{1}{l|}{ACC} & AUC & \multicolumn{1}{l|}{WER} & \multicolumn{1}{l|}{CER} & \multicolumn{1}{l|}{ACC} & AUC \\ \hline
verbatim transcripts & \multicolumn{1}{l|}{--} & \multicolumn{1}{l|}{--} & \multicolumn{1}{l|}{--} & -- & \multicolumn{1}{l|}{--} & \multicolumn{1}{l|}{--} & \multicolumn{1}{l|}{0.826} & 0.873 \\ \hline
\multicolumn{1}{|l|}{Wav2Vec2} &&&&&&&&\\ \hline
\multicolumn{1}{|r|}{base-960h} & \multicolumn{1}{l|}{0.559} & \multicolumn{1}{l|}{0.357} & \multicolumn{1}{l|}{0.787} & 0.911 & \multicolumn{1}{l|}{0.438} & \multicolumn{1}{l|}{0.299} & \multicolumn{1}{l|}{0.830} & 0.908 \\ \hline
\multicolumn{1}{|r|}{large-960h} & \multicolumn{1}{l|}{0.493} & \multicolumn{1}{l|}{0.292} & \multicolumn{1}{l|}{0.830} & 0.897 & \multicolumn{1}{l|}{0.427} & \multicolumn{1}{l|}{0.266} & \multicolumn{1}{l|}{0.787} & 0.855 \\ \hline
\multicolumn{1}{|r|}{large-960h-lv60}  & \multicolumn{1}{l|}{0.443} & \multicolumn{1}{l|}{0.252} & \multicolumn{1}{l|}{0.851} & 0.875 & \multicolumn{1}{l|}{0.364} & \multicolumn{1}{l|}{0.254} & \multicolumn{1}{l|}{0.830} & 0.909 \\ \hline
\multicolumn{1}{|r|}{large-960h-lv60-self}  & \multicolumn{1}{l|}{0.422} & \multicolumn{1}{l|}{0.258} & \multicolumn{1}{l|}{0.787} & 0.891 & \multicolumn{1}{l|}{0.354} & \multicolumn{1}{l|}{0.234} & \multicolumn{1}{l|}{0.809} & 0.879 \\ \hline
\multicolumn{1}{|l|}{HuBERT} &&&&&&&& \\ \hline
\multicolumn{1}{|r|}{large-ls960-ft} & \multicolumn{1}{l|}{0.415} & \multicolumn{1}{l|}{0.228} & \multicolumn{1}{l|}{0.851} & 0.911 & \multicolumn{1}{l|}{0.332} & \multicolumn{1}{l|}{0.210} & \multicolumn{1}{l|}{0.809} & 0.857 \\ \hline
average & \multicolumn{1}{l|}{0.466} & \multicolumn{1}{l|}{0.277} & \multicolumn{1}{l|}{--} & -- & \multicolumn{1}{l|}{0.383} & \multicolumn{1}{l|}{0.253} & \multicolumn{1}{l|}{--} & -- \\ \hline
\end{tabular}

\label{tab:res}
\end{table*}

\paragraph{Both pre-trained and fine-tuned ASR models delivered subpar results on the audio segments from dementia patients.}Pre-trained and fine-tuned ASR models produced significantly less (one-sided Welch's t-test p-value $<$ 0.01) WER/CER on audio segments from healthy controls than dementia patients in the ADReSS test set. Additionally, we observed significantly negative Spearman correlations between WER/CER and MMSE for both pre-trained and fine-tuned ASR models. Higher MMSE (less cognitive impairment) was associated with lower WER/CER and vice versa. As shown in Table~\ref{tab:generated-trans} in Appendix~\ref{apd}, we found that both pre-trained and fine-tuned ASR models generated more empty predictions, more words that were not spelled correctly but sounded very similarly, and more phonetic-level transcripts from audio segments from dementia patients, all of which decreased the performance of ASR models.

\paragraph{Fine-tuning ASR models did not always lead to an improvement the subsequent classification task using the ASR-generated transcripts.}Surprisingly, we found fine-tuning had mixed effects on performance on the downstream task. For example, fine-tuning {Wav2Vec2-base-960h} and {Wav2Vec2-large-960h-lv60-self} increased  downstream classification task accuracy, but the corresponding AUCs dropped. We observed the opposite fine-tuning results with {Wav2Vec2-large-960h-lv60}: downstream task accuracy decreased but the AUC increased. Neither {Wav2Vec2-large-960h} nor {hubert-large-ls960-ft} benefited from fine-tuning: both accuracy and AUC on the downstream task decreased. The lowest WER/CER was obtained by the pre-trained and fine-tuned {hubert-large-ls960-ft} model, however its transcripts did not yield the best downstream task results.

\section{Discussion}

Our findings reveal a paradox: despite the WER/CER of the ASR models evaluated in this study being relatively high compared to their performance on the open domain dataset, the ASR-generated transcripts (both from pre-trained and fine-tuned models) are still able to support acceptable discriminating power of the downstream BERT classification model for differentiating between language produced by cognitively healthy controls and AD patients. Both pre-trained and fine-tuned ASR models perform worse on the audio segments produced by dementia patients, which degrade ASR model performance. Lastly, we find significantly negative correlations between WER/CER and MMSE, suggesting that cognitive impairment results in speech that is more difficult for ASR to recognize, which may limit the applicability of ASR models to transcribe text in the clinical setting.

As we expected, fine-tuning ASR models on an independently collected dataset can improve the performance of ASR models. Our fine-tuning results are consistent with the prior observation \citep{min-etal-2021-evaluating} that augmenting ASR models with additional context may alleviate the impact of noisy transcriptions and audio recordings. HuBERT's improved performance is also consistent with prior work \citep{9688137} indicating that pre-trained HuBERT supports SOTA performance in many downstream tasks, and can generalize well to multiple datasets unrelated to the one on which it was trained. 

Unlike prior works using transcripts or audio signals from the ADReSS/WLS dataset for classification, our approach mediates explainability of the deep neural network (DNN) classifier by providing transcripts generated by SSL ASR models, offering some transparency for healthcare providers to help them understand the diagnostic and other decisions made by DNNs (e.g. by showing which word tokens influenced a prediction of dementia). For example, \citet{guo2021crossing} used professionally transcribed verbatim transcripts of the WLS dataset for data augmentation during BERT fine-tuning and did not use any ASR-generated transcripts. \citet{haulcy2021classifying} employed i-vectors and x-vectors as audio features, which were extracted from audio recordings directly. While i-vectors and x-vectors are efficient acoustic features for DNNs, they usually require extra steps between source audio and the generated transcripts. Unlike using ASR-generated text as an intermediate representation, using i-vectors and x-vectors as input features for a DNN obscures the explainability of DNN predictions, offering limited guidance for further clinical applications.


Our work has several limitations. First, the poor quality of ADReSS audio recordings is very challenging for ASR models, which prevented us for exploring the lower range of WER/CER in an ecologically valid fashion (i.e., without artificially manipulating ASR-generated transcripts). There is a critical need for public availability of high-quality audio recordings of the ``Cookie Theft" picture description task performed by patients with dementia and controls accompanied by neuropsychological evaluation scores. Second, the ADReSS and the WLS data are in American English, and many participants of these two studies are representative of White, non-Hispanic American men and women located at the north part of the United States. Performance with ASR models may differ in populations with different accents \citep{prasad-jyothi-2020-accents}. Third, WER and CER quantify ASR performance, but identifying the nature of the errors concerned is critical to dementia screening and monitoring. Thus further research is needed to study the source(s) of errors and their effects on this task. Lastly, it should be noted that all of the ASR models we tested in this study were pre-trained on read speech, whose nature is very different from the spontaneous speech \citep{8268245, HOWELL1991163}. This may help to explain why WER and CER differ on the ``Cookie Theft" audio recordings, regardless of the audio quality. ASR models that are pre-trained on spontaneous speech would potentially benefit the downstream task. 

\section{Conclusion}

In this work, we investigated the utility of using ASR-generated transcripts in lieu of manual transcripts for discriminating spontaneous speech of dementia patients from that of controls. We found that fine-tuning on a held-out dataset improved the ASR transcription accuracy, but resulted in mixed performance on the subsequent classification task. WER/CER of fine-tuned ASR models were still high in comparison to open-domain datasets. Paradoxically, we found that lower WER/CER did not always boost the subsequent classification performance. Further research is needed to determine the exact nature of this effect and whether it can be systematically leveraged to improve downstream task performance.

\acks{This research was supported by a grant from the National Institute on Aging (AG069792). }
\bibliography{li2022}

\appendix
\section{Supplemental Resources}\label{apd}
\begin{figure}[htbp]
    \centering
    \small
    \includegraphics[scale=0.25]{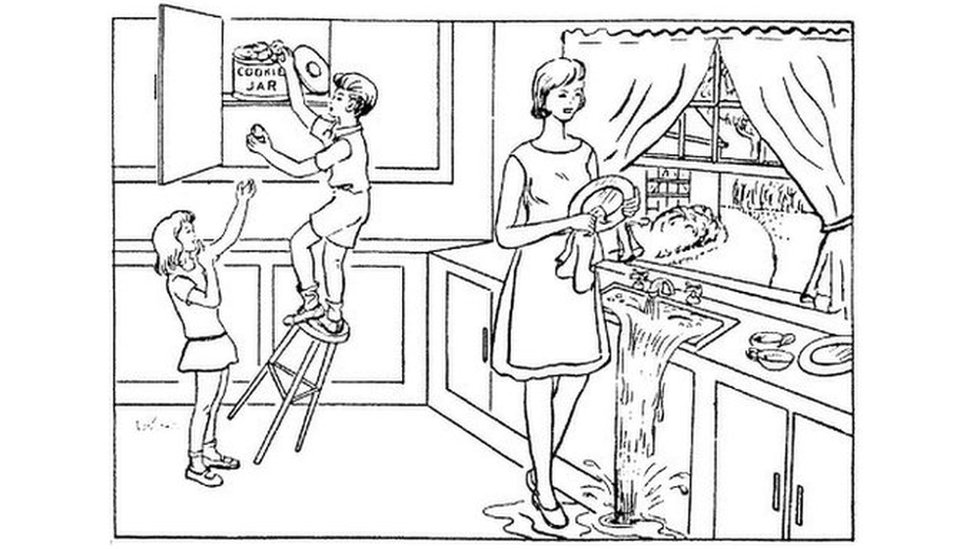}
    \caption{``Cookie Theft'' picture stimulus}
    \label{fig:cookie-theft}
\end{figure}

\begin{table}[htbp]
\small
\centering
\caption{Variants of SSL ASR models structure and performance on LibriSpeech's ``clean'' and ``other'' test data.}
\begin{tabular}{|p{3cm}|p{1cm}|ll|}
\hline
\multirow{2}{*}{\textbf{Model}} & \multirow{2}{1cm}{\textbf{Blocks}} & \multicolumn{2}{l|}{\textbf{WER \%}}    \\ \cline{3-4} 
                  &                   & \multicolumn{1}{l|}{clean} & other \\ \hline
Wav2Vec2-base-960h & 12  & \multicolumn{1}{l|}{3.4} & 8.6 \\ \hline
Wav2Vec2-large-960h & 24 & \multicolumn{1}{l|}{2.8} & 6.3  \\ \hline
Wav2Vec2-large-960h-lv60 & 24  & \multicolumn{1}{l|}{2.2} & 4.6 \\ \hline
Wav2Vec2large-960h-lv60-self & 24 & \multicolumn{1}{l|}{1.9} & 3.9  \\ \hline
hubert \newline large-ls960-ft & 16& \multicolumn{1}{l|}{1.9} & 3.3 \\ \hline
\end{tabular}

\label{tab:models}
\end{table}

\begin{equation}
    \begin{aligned}
    W/CER & = \frac{S+D+I}{N} = \frac{S+D+I}{S+D+C}\\
    \end{aligned}
    \label{eq:wer}
\end{equation}

where $S$ is the number of substitutions, $D$ is the number of deletions, $I$ is the number of insertions, $C$ is the number of correct words/characters, and $N$ is the number of words/characters in the reference. The lower WER/CER is, the better the model performs.

\paragraph{Performance of ASR models on the WLS test set}On the WLS test set, the pre-trained {Wav2Vec2-base-960h}, {Wav2Vec2-large-960h}, {Wav2Vec2-large-960h-lv60}, {Wav2Vec2-large-960h-lv60-self}, and {hubert-large-960h-ft} achieved WER at 0.541, 0.471, 0.412, 0.390, and 0.332, CER at 0.318, 0.269, 0.240, 0.235, and 0.210, respectively. The fine-tuned {Wav2Vec2-base-960h}, {Wav2Vec2-large-960h}, {Wav2Vec2-large-960h-lv60}, {Wav2Vec2-large-960h-lv60-self}, and {hubert-large-960h-ft} achieved WER at 0.366, 0.358, 0.277, 0.264, and 0.266, CER at 0.227, 0.210, 0.176, 0.159, and 0.153.

\begin{table}[htbp]
\small
\centering
\caption{Comparison of verbatim, pre-trained, and fine-tuned ASR-generated transcripts and the subsequent classification prediction for ADReSS ID 148-0, a dementia patient whose MMSE is 10. }
\label{tab:generated-trans}
\begin{minipage}{0.8\textwidth}
\begin{tabular}{|p{2.5cm}|p{9cm}|l|}
\hline
 \textbf{Model}& \textbf{Transcripts} & \textbf{Prediction} \\\hline
Verbatim & the sink is running over. the girl's reaching for a cookie. the mom is drying a dish. cup and saucers there. yeah that's about all that. i didn't hit did this and then i did that. well yeah here's some outside the window. a garden i guess  & healthy\\\hline
pre-trained Wav2Vec2-base-960h & saint strunning oer.  . try. jars.  . thithiis. wile ye i hear some outside wonde. ardis & dementia \\\hline
fine-tuned Wav2Vec2-base-960h & the sinks running o.  . ddrying a dishes. ac.  . i i.  .  & dementia\\\hline
pre-trained Wav2Vec2-large-960h & the sing is running over. cos ce flo coky. o gina dish. me capin sawses.  . an an bism. well ye i hersome outside wenda. gardngats & healthy \\\hline
fine-tuned Wav2Vec2-large-960h & the sink is plioting over. reaching for a cookie. woman drying the dish. cup and saucers.  . i tan o. well  a heron outside window. & healthy\\\hline
pre-trained Wav2Vec2-large-960h-lv60 & the sing is running over. cos ce flo coky. o gina dish. me capin sawses.  . an an bism. well ye i hersome outside wenda. gardngats & healthy \\\hline
fine-tuned Wav2Vec2-large-960h-lv60 & the sink is plioting over. reaching for a cookie. woman drying the dish. cup and saucers.  . i tan o. well  a heron outside window. & healthy\\\hline
pre-trained Wav2Vec2-large-960h-lv60-self & the same running over. t four fift. dry landit. captain sar seems t. teno. in dimplest. er s outside window. garden against & dementia \\\hline
fine-tuned Wav2Vec2-large-960h-lv60-self & thesink is runing oer. fora c. m. u. u. i didn't hav sid l. ss. u & dementia\\\hline
pre-trained hubert-large-960h-ft & the sant is running over. t asociation for a party. o drying a dish. captain sawsewas. e a it  man always o. tan itentes. well ye hear some outside winda. ou garden against & dementia\\\hline
fine-tuned hubert-large-960h-ft & the sink is running over. ling for a cookie. drying a dish. cup and saucer. utaa. i didn't. well  har someoutsidewindow. ua & dementia\\\hline

\end{tabular}
\end{minipage}
\end{table}

\end{document}